\documentclass[twocolumn,showpacs,amsmath,amssymb,superscriptaddress,prl]{revtex4-1}

\usepackage{graphicx}
\usepackage{dcolumn}
\usepackage{bm}
\usepackage[abs]{overpic}
\usepackage{here}
\usepackage{color}
\usepackage{url}
\usepackage{accents}
\def\up{\uparrow}
\def\down{\downarrow }

\newcommand{\exn}{\langle n \rangle}

\begin{document}


\title{
Self-Learning Monte Carlo Method: Continuous-Time Algorithm
}

\author{Yuki Nagai}
\affiliation{CCSE, Japan  Atomic Energy Agency, 178-4-4, Wakashiba, Kashiwa, Chiba, 277-0871, Japan}
\affiliation{Department of physics, Massachusetts Institute of Technology, Cambridge, MA 02139, USA}

\author{Huitao Shen}
\affiliation{Department of physics, Massachusetts Institute of Technology, Cambridge, MA 02139, USA}

\author{Yang Qi}
\affiliation{Department of physics, Massachusetts Institute of Technology, Cambridge, MA 02139, USA}

\author{Junwei Liu}
\affiliation{Department of physics, Massachusetts Institute of Technology, Cambridge, MA 02139, USA}

\author{Liang Fu}
\affiliation{Department of physics, Massachusetts Institute of Technology, Cambridge, MA 02139, USA}

\date{\today}

\begin{abstract}
The recently-introduced self-learning Monte Carlo method is a general-purpose numerical method that speeds up Monte Carlo simulations by training an effective model to propose uncorrelated configurations in the Markov chain.
We implement this method in the framework of continuous time Monte Carlo method with auxiliary field in quantum impurity models.
We introduce and train a diagram generating function (DGF) to model the probability distribution of auxiliary field configurations in continuous imaginary time, at all orders of diagrammatic expansion. By using DGF to propose global moves in configuration space, we show that the self-learning continuous-time Monte Carlo method can significantly reduce the computational complexity of the simulation.
\end{abstract}

\maketitle

Quantum Monte Carlo (QMC) is an unbiased numerical method for studying quantum many-body systems. A standard QMC scheme for interacting fermion systems is the determinantal QMC method \cite{Blankenbecler,Hirsch1985,Hirsch1986,White}. This method uses (1) the Hubbard-Stratonovich transformation to decompose the two-body fermion interaction, and (2) the Suzuki-Trotter decomposition of the partition function to  discretize the imaginary time interval into a large number of time slices.  Monte Carlo sampling is performed in the space of auxiliary Hubbard-Stratonovich fields.
Recently, a continuous-time modification of the fermionic QMC algorithm was developed \cite{Rubtsov,Werner,GullRMP, GullEPL}. In this algorithm, the partition function is expanded in the powers of interaction, and the Monte Carlo simulation is performed by the stochastic sampling of the diagrammatic expansion of interaction terms. Both the number and position of interaction terms on the imaginary time interval change constantly during the simulation. For both determinantal and continuous-time QMC methods, to compute the weight of each configuration requires integrating out the fermions. This is very time-consuming and in practice limits the size of fermion systems in QMC studies.

Recently, we introduced a new general method, dubbed self-learning Monte Carlo (SLMC), which speeds up the MC simulation by designing and training a model to propose efficient global updates \cite{Junwei201610,Junwei201611,Xu}. The philosophy behind SLMC is ``first learn, then earn''. In the learning stage, trial simulations 
are performed to generate a large set of configurations and their weights. These data are then used to train an effective model $H_{\rm eff}$, whose Boltzmann weight $e^{-\beta H_{\rm eff}}$ fits the probability distribution of the original problem. Next, in the actual simulation, $H_{\rm eff}$ is used as a guide to propose highly efficient global moves in configuration space. Importantly, the acceptance probability of such global update is set by the detailed balance condition of the original Hamiltonian. This ensures the MC simulation is statistically exact.

SLMC method is ideally suited for QMC simulation of fermion systems. In the determinantal QMC method, the weight of an auxiliary field configuration $\phi(x)$  is computed by integrating out fermions, which is numerically expensive. In contrast, the effective model $H_{\rm eff}[\phi(x)]$ is an explicit functional of $\phi(x)$, and its Boltzmann weight can be computed fast. Therefore, the SLMC method has a far less computational cost than the original method, leading to a  dramatic speedup as we demonstrated in previous works \cite{Junwei201611}.

In this work, we extend the SLMC to continuous-time quantum Monte Carlo algorithms for fermion systems. Based on theoretical analysis and numerical study, we  demonstrate that our continuous-time SLMC reduces the computational complexity of the simulation in the low-temperature or strong coupling regime, where the autocorrelation time in the standard method becomes large.
The key ingredient of our method is an effective model for the diagrammatic interaction expansion in continuous time, which we term ``diagram generating function'' (DGF).
The form of DGF is constrained by the symmetry of the Hamiltonian under study.  The parameters in DGF are trained and optimized in the learning stage of SLMC.
As an example, we implement SLMC to simulate the single impurity Anderson model \cite{Anderson}, using the continuous-time auxiliary-field (CT-AUX) method \cite{GullEPL,GullRMP,WernerPRB,GullPRB2011}. The DGF for this model is found to take a remarkably simple form, and reproduce with very high accuracy the exact distribution of auxiliary fields in continuous time, to all orders of the diagrammatic expansion. We demonstrate the speedup of SLMC in comparison to the standard CT-AUX, and find the acceleration ratio increases with the average expansion order.

The paper is organized as follows: We first briefly review the CT-AUX algorithm in the Anderson impurity model, after which we give a detailed introduction to the self-learning CT-AUX algorithm, and discuss the physical ideas behind the DGF. Then we show the performance of our new algorithm on the Anderson model. Finally we analyze the complexity of the algorithm.
The technical details are shown in the supplemental materials\cite{Sup}.

While this work is being performed, a related work \cite{LHuang} also extending SLMC \cite{Junwei201610,Junwei201611,LHuangPRB} to continuous time domain appeared. Unlike ours, that work uses interaction expansion without auxiliary field, and does not analyze the computational complexity of continuous-time SLMC to demonstrate its speedup.

\paragraph{CT-AUX Method} The Hamiltonian of the single impurity Anderson model is written as the combination of a free fermion part and an interaction part \cite{GullEPL}
\begin{align}
H &= H_0+H_1 \label{eq:h}\\
H_{0} &= -(\mu - U/2) (n_{\up} + n_{\down}) + \sum_{\sigma,p} (V c_{\sigma}^{\dagger} a_{p,\sigma} + h.c.) \nonumber \\
&+ \sum_{\sigma,p} \epsilon_{p} a_{p,\sigma}^{\dagger} a_{p,\sigma} +K/\beta, \\
H_{1} &= U(n_{\up} n_{\down} - (n_{\up} + n_{\down})/2) - K/\beta,
\end{align}
where $ \sigma=\uparrow,\downarrow $, $c_{\sigma}^{\dagger}$ and $a_{p,\sigma}^{\dagger}$ are the fermion creation operators for an impurity electron with spin $\sigma$, and that for a bath electron with spin $\sigma$ and momentum $p$, respectively. $n_{\sigma} = c_{\sigma}^{\dagger} c_{\sigma}$ is the fermion number operator. $\beta=1/T$ is the inverse temperature. $ K $ is an arbitrary chosen parameter controls the coupling strength of the auxiliary field and the average expansion order, which we will see below.

In the CT-AUX method, the density-density interaction in $H_{1}$ is decoupled by an auxiliary Ising field $ s $ as
\begin{align}
H_{1} &= -\left(\frac{K}{2 \beta} \right) \sum_{s=\pm 1} e^{\gamma s (n_{\up} - n_{\down})}.
\end{align}
$ \gamma $ is the coupling strength between the fermion density and the auxiliary field, and is determined by $\cosh (\gamma) \equiv 1+(\beta U)/(2K)$. The partition function is expanded as
\begin{align}
 \frac{Z}{Z_{0}} &= {\rm Tr} \: \left[
e^{- \beta H_{0}}
T_{\tau} e^{
-\int_{0}^{\beta} d \tau H_{1}(\tau)
 }
\right],\nonumber \\
&= \sum_{n=0} \int_{0}^{\beta} d\tau_{1} \cdots \int_{\tau_{n-1}}^{\beta} d\tau_{n} \left(
\frac{K}{2\beta}
 \right)^{n} \frac{Z_{n}(\{s_{i},\tau_{i} \})}{Z_{0}}. \label{eq:oriz}
\end{align}
Here
\begin{eqnarray}
Z_{n}(\{s_{i},\tau_{i} \})/Z_{0} &\equiv& \prod_{\sigma=\up,\down} {\rm det} \: N_{\sigma}^{-1}(\{s_{i},\tau_{i} \}), \label{eq:weight} \nonumber\\
N_{\sigma}^{-1}(\{s_{i},\tau_{i} \}) &\equiv&  e^{V_{\sigma} \{s_{i} \}} - G_{0\sigma}^{\{ \tau_{i}\}} (e^{V_{\sigma} \{s_{i} \}} -1)
\end{eqnarray}
where $Z_{0} \equiv {\rm Tr} \: e^{- \beta H_{0}}$, and
$e^{V_{\sigma} \{s_{i} \}} \equiv {\rm diag} \left(e^{\gamma (-1)^{\sigma} s_{1}},\cdots,e^{\gamma (-1)^{\sigma} s_{n}}\right)$ with
the notations $(-1)^{\up} \equiv 1$, $(-1)^{\down} \equiv -1$, $(G_{0\sigma}^{\{ \tau_{i}\}})_{ij} = g_{\sigma}(\tau_{i} - \tau_{j})$ for $i \ne j$, and $(G_{0\sigma}^{\{ \tau_{i}\}})_{ii} = g_{\sigma}(0^{+})$.  $g_{\sigma}(\tau)>0$ is the free fermion Green's function at the impurity site. The configuration space for the MC sampling is hence the collection of all the possible auxiliary spin configurations on the imaginary time interval and at all possible expansion orders $n=0,1,...$, $c = \{ \{\tau_{1},s_{1}\} \cdots \{\tau_{n},s_{n} \} \}$ where $0\leq \tau_1<\tau_2<\ldots<\tau_n < \beta$ and $s_{i} = \uparrow,\downarrow$.

The corresponding weight $ w_c $ is given by Eq.~\eqref{eq:weight}. Then a random walk $c_{1} \rightarrow c_{2} \rightarrow c_{3} \rightarrow \cdots$ in configuration space is implemented usually by inserting/removing random spins at random imaginary times.

\paragraph{Self-learning CT-AUX}
In this section, we describe the self-learning continuous-time auxiliary-field method. Like other SLMC methods, it consists of two parts: (1) learn an effective model or DGF that approximates the probability distribution of auxiliary spins in imaginary time interval $\{ \{\tau_{1},s_{1}\} \cdots \{\tau_{n},s_{n} \} \}$, and (2) propose a global move by executing a sequence of local updates in the effective model \cite{Junwei201611}.

Since the number of auxiliary spins changes constantly with the expansion order $n$ in the sampling process, one may expect that to reproduce the entire probability distribution at all orders requires a highly sophisticated model with a huge number of parameters. On the contrary, we introduce a DGF of a remarkably simple form which fits the probability distribution very accurately
\begin{align}
Z_{n}(\{s_{i},\tau_{i} \} )/Z_{0} \simeq e^{-\beta H^{\rm eff}_n(\{s_{i},\tau_{i} \})},\label{eq:effw}
\end{align}
\begin{align}
-\beta H^{\rm eff}_{n}(\{s_{i},\tau_{i} \}) &\equiv \frac{1}{n}\sum_{i,j} J(\tau_{i} - \tau_{j} )s_{i} s_{j} +\frac{1}{n}\sum_{i,j} L(\tau_{i} - \tau_{j}) \nonumber \\
&+ f(n). \label{eq:effh}
\end{align}
Several features of $H^{\rm eff}_n$ deserve attention:

(i) DGF serves as an approximation to $ Z_n $ in the weak coupling expansion as is indicated in Eq.~\eqref{eq:effw}, whose functional form could be obtained exactly if one could integrate out fermion degrees of freedom exactly. This is indeed what is done in the original CT-AUX algorithm. Here in SLMC, the DGF is instead constructed by series expansion and symmetry analysis. To be specific, Eq.~\eqref{eq:effh} is the spin-spin interactions satisfying the spin-flip symmetry $s_i \rightarrow -s_i$ up to two-body terms. Since the performance of the DGF is already good enough at this stage, we did not include fourth-order interactions that are proportional to $ s_is_js_ks_l $. 

(ii) The interaction terms $J(\tau)$ and $L(\tau)$ are in principle allowed to be different functions of $\tau$ at different expansion orders $n$, which would result in vastly more parameters.
Here this predicament is avoided by choosing the same functions to all expansion orders. 

(iii) The expansion-order dependent factor $1/n$ in Eq.~\eqref{eq:effw} is crucial. It can be justified by considering the atomic limit $V = 0$, where the interaction term $H_1(\tau)\equiv H_1$ in Eq.~(\ref{eq:oriz}) becomes  independent of $\tau_i$, and hence $Z_n \propto {\rm Tr} (H_1^n)$. For large $n$, ${\rm Tr} (H_1^n) \simeq \epsilon_0^n$ is dominated by the contribution from the largest eigenvalue $\epsilon_0$, hence $ \ln Z_{n}/Z_{0}$ increases linearly with $n$.
On the other hand, $H^{\rm eff}_n$ in Eq.~(\ref{eq:effh}) includes a summation of $n^2$ pairwise interactions at pairs of imaginary time instances $(\tau_i, \tau_j)$.  Therefore we must include the factor $1/n$ to match the two results. 

As we will show later, this simple DGF performs remarkably well.

The training procedure goes as follows. Given a set of configurations $ \{c_i\} $ taken from the Markov chain of a MC simulation, we minimize the mean square error $ \left(\ln Z^{\rm eff}_{n}-\ln Z_{n}/Z_0\right) $ on this training set by varying the functional form of $J(\tau) $, $ L(\tau) $ and $ f(n) $. In practice, we use Chebyshev polynomials $T_{m}(x) = \cos (m \arccos(x))$ to expand functions $J$ and $L$, $J(\tau) \equiv \sum_{m=0}^{m_{c,J}} a_{m} T_{2m}(x(|\tau|))$ and $L(\tau) \equiv \sum_{m=0}^{m_{c,L}} b_{m} T_{2m}(x(|\tau|))$ with $x(\tau) \equiv 2\tau/\beta -1$ \cite{Wolf,Braun,Covaci,NagaiCheb,Sota}, and use power series to expand the function $f$, $f(n) = \sum_{k=0}^{m_{c,f}} c_{k} n^{k}$. Here $ m_{c,J} $, $ m_{c,L}  $ and $ m_{c,f}  $ are the truncation orders for the respective functions. The rationale behind the choice of basis functions is that the Chebyshev polynomial is close to the minimax polynomial minimizing the maximum error in the approximation. In other word, the Chebyshev polynomials approximate the original function uniformly \cite{note}. In the simulation, we always increase the truncation order until the results converge. The total number of training parameters is thus $ m_{c} \equiv m_{c,J}+m_{c,L}+m_{c,f}+3 $ (summation starts from 0) \cite{note2}. Since the DGF $H^{\rm eff}_n$ is a linear function of these parameters, they can be trained simply with a linear regression \cite{stab}.
We have also exploited iterative training procedure to improve the efficiency \cite{Junwei201610}, whereby Monte Carlo configurations and weights generated by the self-learning algorithm are used as training data to further train the DGF. This procedure can be iterated until the DGF reproduces exact probability distribution sufficiently well. We note that training the effective model can be regarded as supervised learning in a broader context of machine learning, which recently has many fruitful applications in physics \cite{Carrasquilla,Carleo,Hu,Deng,TanakaTomiya,Fujita,YiZhang,JChen,YHuang,ZiCai,SJohann,ENieuwenberg,Zdeborova}.

\begin{figure}[t]
\centering
\includegraphics[width=0.9\columnwidth]{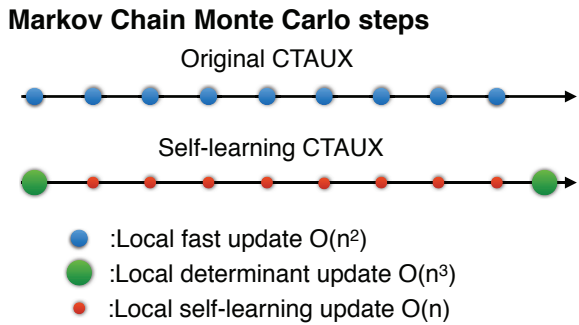}
\caption{
(Color online) Schematic figure for the Markov chains in the original and self-learning continuous-time Monte Carlo methods to obtain an uncorrelated configuration. $n$ denotes the average expansion order that determines the size of the matrix $N_{\sigma}(\{s_{i},\tau_{i} \})$, and further determines the complexity of the simulation. See the last section of the paper for a detailed discussion. }
\label{fig:fig5}
\end{figure}

After completing the training process, we use the trained DGF to propose highly-efficient global moves on the Markov chain in actual simulations. Here we adopt the general procedure of cumulative update introduced in Ref.~\cite{Junwei201611}.
Fig.~\ref{fig:fig5} illustrates how self-learning CT-AUX proposes global moves, in comparison with the original CT-AUX method. Starting from a configuration $ c_i $, we perform a sufficiently large number (denoted by $M_{\rm eff}$) of local updates by inserting/removing random spins at random imaginary times based on the weights of the DGF, until reaching a  configuration $ c_j $ that is sufficiently uncorrelated with $c_i$. The global move $c_i \rightarrow c_j $ is then proposed, and its acceptance rate $p$ is calculated from the exact weight of the original model, $ p= \min\{1,(w_{c_j}w^{\rm eff}_{c_i})/(w_{c_i}w^{\rm eff}_{c_j})\} $, where $ w_{c_i} $ and $ w^{\rm eff}_{c_i} $ are weights of configuration $ c_i $ computed from the original model Eq.~\eqref{eq:oriz} and effective model Eq.~\eqref{eq:effw} respectively. As shown previously \cite{Junwei201611}, this cumulative update procedure fulfills  the ergodicity condition and obeys the detailed balance principle. Since computing the weight of DGF is much faster than computing the fermion determinant in the original method, our method significantly reduces the computational cost of the simulation.
A detailed discussion on the choice of the cumulative update length $ M_{\rm eff} $ and the computational complexity of self-learning CT-AUX method is presented in the last section of this work.

\begin{figure}[t]
\centering
\includegraphics[width=0.9\columnwidth]{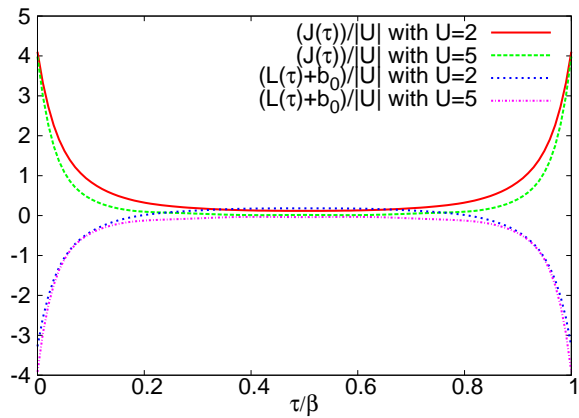}
\caption{(Color online)  Effective interactions for different $ U $ with $\beta = 10$, $V = 1$, and $K= 1$. }
\label{fig:fig2}
\end{figure}

\paragraph{Performance on Anderson Model}
Now we are ready to show the performance of self-learning CT-AUX on the single impurity Anderson model. We consider a bath with a semicircular density of states $\rho_{0}(\epsilon) = (2/(\pi D) \sqrt{1-(\epsilon/D)^{2}})$ and set the half bandwidth $D = 1$ as the energy unit. The chemical potential is set to be  $\mu = U/2$ to maintain a half-filling.

\begin{figure}[t]
\centering
\includegraphics[width=\columnwidth]{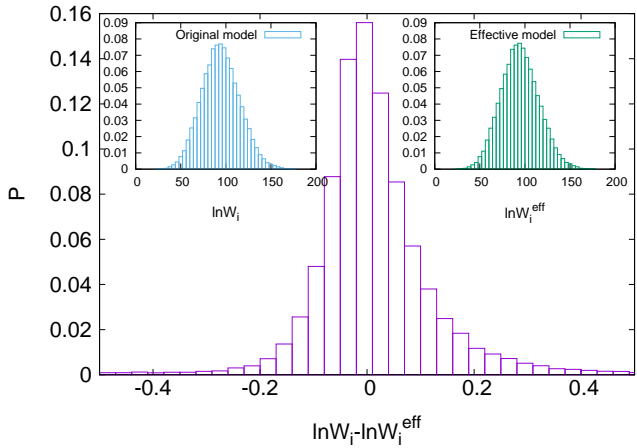}
\caption{
(Color online)  For $5 \times 10^{4}$ independent configurations on the Markov chain of the original CT-AUX, histogram is the distribution of the difference $\ln W_{i}-\ln W_{i}^{\rm eff}$. The upper-left and upper-right insets are distributions of $\ln W_{i}$ and $\ln W_{i}^{\rm eff}$, respectively. Here $U=5$, $V=1$, $\beta = 10$, and $K = 1$. }
\label{fig:fig1}
\end{figure}

In the simulation, we use $ 5\times 10^4 $ configurations as the training data set. Throughout the parameter regime in our calculations, a total of 30 training parameters ($m_{c,J}=m_{c,L}=12$, $ m_{c,f}=3$) is enough to guarantee the convergence of the DGF.
After training, we obtain the interaction functions $J(\tau)$ and $L(\tau)$ in the DGF (\ref{eq:effh}), as shown in Fig.~\ref{fig:fig2}. They become more localized at $\tau=0$ and $\beta$ with increasing $U$. To evaluate the accuracy of the DGF,
we plot in Fig.~\ref{fig:fig1} the distribution of the weights of the DGF and those of the original model exactly computed. The two distributions look very similar. To   quantitatively measure the goodness of fit, we evaluate the quantity $R^2\in [0,1]$ which is introduced as the ``score'' of self-learning Monte Carlo method in general \cite{Junwei201611}. Here, we find the DGF for the Anderson impurity model (with $U=5$, $\beta=10$, $V=1$, and $K=1$) has a score of $ R^2=99.9 \%$. Thanks to the  success of our DGF,  a global move proposed by cumulative update between two uncorrelated configurations has a very high average acceptance rate around $0.68$.

\begin{figure}[t]
\centering
\includegraphics[width=0.95\columnwidth]{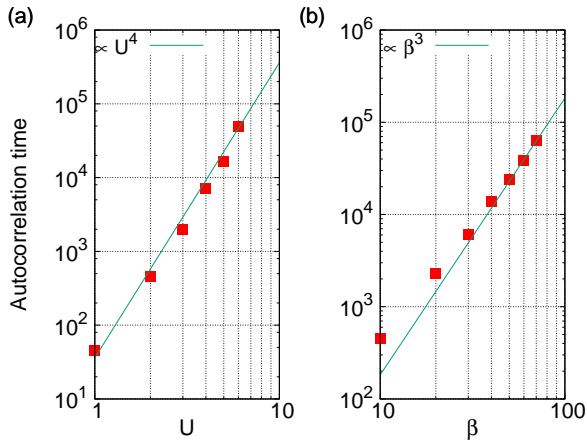}
\caption{
(Color online) (Left panel) $U$-dependence of the autocorrelation time of the original CT-AUX with $\beta = 10$.
(Right panel)  $\beta$- dependence of the autocorrelation time with $U = 2$. The other related parameters are $V = 1$ and $K= 1$. In both of the figures, the unit time is a local update (inserting/removing a auxiliary spin) in the original CT-AUX method.
}
\label{fig:fig4}
\end{figure}

\begin{figure}
\centering
\includegraphics[width=.95\columnwidth]{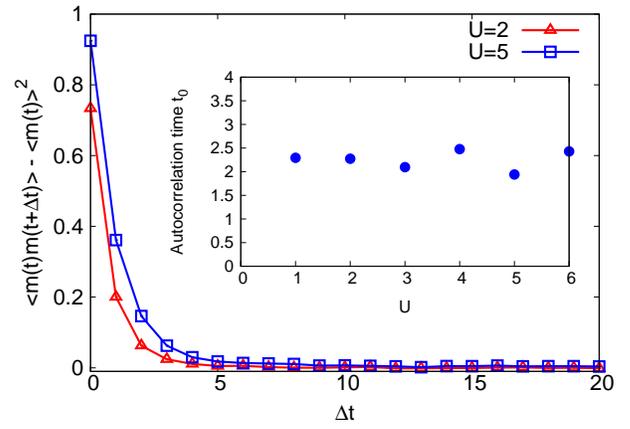}
\caption{
(Color online)  Autocorrelation function of the auxiliary-spin magnetization for a system with $\beta = 10$, $V = 1$, and $K= 1$. Unit time is defined in the main text.
(Inset) $U$ dependence of the autocorrelation time in the self-learning CT-AUX.
We set the number of local updates on DGF to be $M_{\rm eff} = 2\times 10^3$ ($U=1,2,3,4$) and $M_{\rm eff} = 5\times 10^4$ ($U=5,6$).
}
\label{fig:fig3}
\end{figure}

To demonstrate the speedup of self-learning CT-AUX method, we compute the autocorrelation function of the auxiliary spin polarization defined by $m \equiv (1/n) \sum_{i=1}^{n} s_{i}$. Fig.~\ref{fig:fig4} shows the autocorrelation time of the original CT-AUX method, defined in terms of the number of local updates. It is clear that the autocorrelation time increases rapidly with $ \beta $ and $ U $, rendering the algorithm inefficient at low temperature and in the strong coupling regime. In contrast, the performance of the self-learning CT-AUX method is shown in Fig.~\ref{fig:fig3}. The autocorrelation function decays rapidly with the number of global moves proposed by the DGF. This is because (1) a single global move is the cumulative outcome of $M_{\rm eff}$ local updates, where $M_{\rm eff}$ is taken to be so large that the proposed configuration is sufficiently uncorrelated from the current one, and (2) the average acceptance rate for such global moves are high enough --- greater than $ 0.6 $ for all the data points in Fig.~\ref{fig:fig3}. The inset shows the $U$ dependence of the autocorrelation time $t_{0}$, which is estimated from the initial slope of the autocorrelation function $\langle m(t) m(t+\Delta t) \rangle \sim e^{-\Delta t/t_{0}}$. 
It is worth noting that with increasing $ M_{\rm eff} $, the autocorrelation time of our self-learning algorithm saturates to a small value even for very large  $ U $.

\paragraph{Computational Complexity} Finally, we discuss the actual calculation cost of the self-learning CT-AUX method. Fig.~\ref{fig:fig5} shows schematically the Markov chains to obtain two uncorrelated configurations. Roughly speaking, self-learning CT-AUX is faster than the original CT-AUX because the computational cost of each local move in the Markov chain is smaller than that in the CT-AUX. A detailed analysis is given as follows. In order to compare the two methods on a equal footing, we consider the cost to obtain an uncorrelated configuration from a given one. In this way, the two methods give the same error bar for the measured observables. The cost for inserting/removing a vertex with the use of fast updates is ${\cal O}(\exn^{2})$ in the original CT-AUX simulation \cite{GullEPL}. $\exn$ is the average expansion order that determines the size of the matrix $N_{\sigma}(\{s_{i},\tau_{i} \})$. To obtain an uncorrelated configuration, $\tau_{\rm ori}$ such local updates are needed. (This is actually the definition of autocorrelation time in the original method. ) Thus, the total cost is ${\cal O}(\exn^{2} \tau_{\rm ori})$. On the other hand, the cost for inserting/removing a vertex is ${\cal O}(m_{c} \exn)$ in the effective model. Recall $ m_c $ is the number of the training parameters in the DGF. The scaling of $ \exn $ is different from that in the original CT-AUX because the weight of DGF is computed directly without calculating the fermion determinant. The number of the local updates using DGF $M_{\rm eff}$ should be $\tau_{\rm ori}$ in order to obtain an uncorrelated configuration. And we need one more calculation of the determinant to decide the weight of the proposed global move, whose computational cost is ${\cal O}(\exn^{3})$. Note that the global move is not always accepted, there is additional $ \tau_{\rm SL} $ overhead, which is the autocorrelation time measured in Fig.~\ref{fig:fig3}. Thus the total calculation cost of the self-learning algorithm is ${\cal O}\left((\exn^{3} + m_c\exn \tau_{\rm ori}\right) \tau_{\rm SL})$. Since $ \exn\sim\beta U  $ \cite{GullEPL} and the autocorrelation time $\tau_{\rm ori}$ is approximately proportional to $U^{4} \beta^{3}$ as shown in Fig.~\ref{fig:fig4}, the second term in the bracket dominants. This is indeed the case shown in the inset in Fig.~\ref{fig:fig3}. In fact, in our computation $ \exn $ is less than 30 while the $ \tau_{\rm ori} $ can be up to of order $ 10^6 $. In this way, the actual speed-up ratio $t_{s}$ is expressed by
\begin{align}
t_{s} &\sim \frac{\exn}{m_c\tau_{\rm SL}}.
\end{align}
As long as the DGF is good enough, $ \tau_{\rm SL} $ is ${\cal O}(1)$. Since $ m_c $ hardly scales with $ U $ and $ \beta $, the self-learning CT-AUX method is generally faster than the original CT-AUX especially in the low temperature and strong coupling regime when $ \exn \sim \beta U $ is large.

\paragraph{Conclusion} We developed the continuous-time version of the SLMC with auxiliary field, which trains an effective model (DGF) to propose new uncorrelated configurations in the Markov chain, with high acceptance rate.
The DGF for Anderson impurity model is found to take a remarkably simple form, and reproduce very well the exact distribution of auxiliary fields in continuous time to all orders of the diagrammatic expansion.
Our method reduces the computational complexity of the simulation in the low-temperature or strong coupling regime, where the autocorrelation time in the standard method becomes large.

Our self-learning CT-AUX method have many potential applications. It can be used as an impurity solver for dynamical mean-field theory, and is ideal for  studying systems near the critical point \cite{NagaiDMFT,Hoshino,HoshinoPRB}, where standard methods suffer from severe critical slowing down. Our method can also be generalized straightforwardly to fermion lattice models.  

\paragraph{Acknowledgment} The calculations were performed by the supercomputing system SGI ICE X at the Japan Atomic Energy Agency. The work at MIT was supported by DOE Office of Basic Energy Sciences, Division of Materials Sciences and Engineering under Award DE-SC0010526. YN was supported by JSPS KAKENHI Grant Number 26800197, the ''Topological Materials Science'' (No. JP16H00995) KAKENHI on Innovative Areas from JSPS of Japan. HS is supported by MIT Alumni Fellowship Fund For Physics.

\clearpage
\newpage
\widetext
\onecolumngrid

\setcounter{equation}{0}
\renewcommand{\thefigure}{S\arabic{figure}} 

\setcounter{figure}{0}

\renewcommand{\thesection}{S\arabic{section}.} 
\renewcommand{\theequation}{S\arabic{equation}} 
\renewcommand{\thetable}{S\arabic{table}} 
\begin{flushleft} 
{\Large {\bf Supplemental material}}
\end{flushleft} 

\section{Self learning updates}
The probability of moving from a configuration $c_{i}$ to a configuration $c_{j}$ can be split into the probability of proposing the move and the probability of accepting it, $ p(c_{i} \rightarrow c_{j}) = p^{\rm prop}(c_{i} \rightarrow c_{j}) p^{\rm acc}(c_{i} \rightarrow c_{j}) $. Then the detailed balance principle implies
\begin{align}
\frac{p^{\rm acc}(c_{i} \rightarrow c_{j})}{p^{\rm acc}(c_{j} \rightarrow c_{i})} &=
\frac{p^{\rm prop}(c_{j} \rightarrow c_{i})}{p^{\rm prop}(c_{i} \rightarrow c_{j})}
\frac{w_{c_{j}}}{w_{c_{i}}}. \label{eq:acc}
\end{align}
In the self-learning CT-AUX, the new configuration $c_{j}$ is proposed based on the effective weight $w_{c_j}^{\rm eff}$. The probability to propose the move $p^{\rm prop}(c_{j} \rightarrow c_{i})/p^{\rm prop}(c_{i} \rightarrow c_{j}) = w_{c_{i}}^{\rm eff}/w_{c_{j}}^{\rm eff}$. Combined with Eq.~\eqref{eq:acc}, we obtain the desired acceptance rate.
This result can be understood intuitively in the limit that the effective weight $w_{c}^{\rm eff}$ is equal to the original weight $w_{c}$ for all configurations. Then we are as if doing the MC update on exactly the original model. Therefore the ``global update'' from configuration $c_{i}$ to configuration $c_{j}$ should always be accepted, i.e., $p^{\rm acc}(c_{i} \rightarrow c_{j})/p^{\rm acc}(c_{j} \rightarrow c_{i}) = w_{c_{i}}^{\rm eff} w_{c_{j}}/(w_{c_{j}}^{\rm eff} w_{c_{i}}) = 1$.

\section{Comparison between the original and effective weights}
\begin{figure}[b]
\begin{center}
     \begin{tabular}{p{ 0.5 \columnwidth}} 
      \resizebox{0.5\columnwidth}{!}{\includegraphics{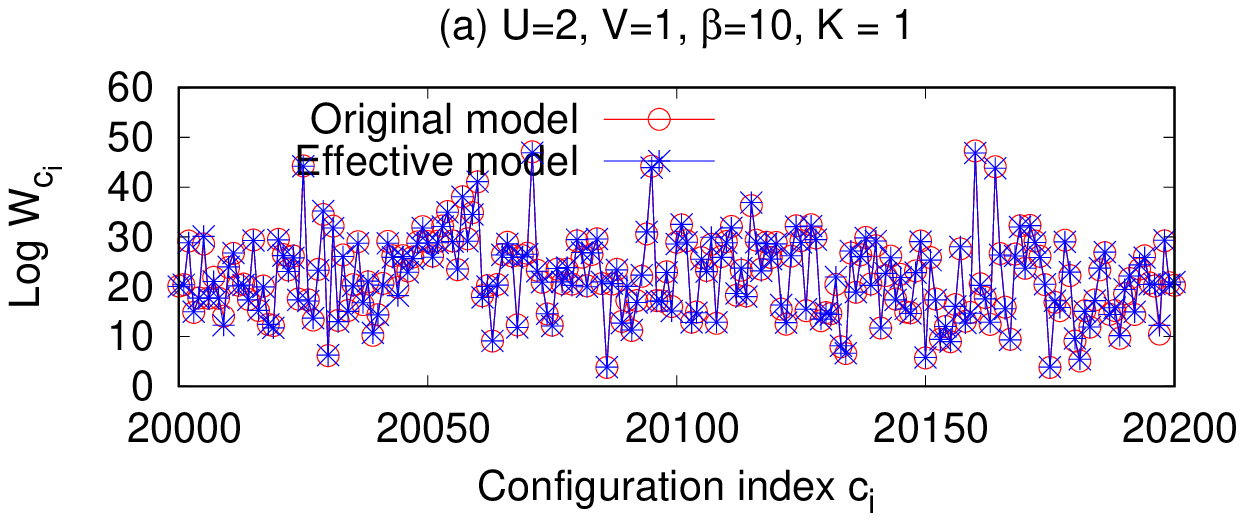}} 
      \\ 
      \resizebox{0.5 \columnwidth}{!}{\includegraphics{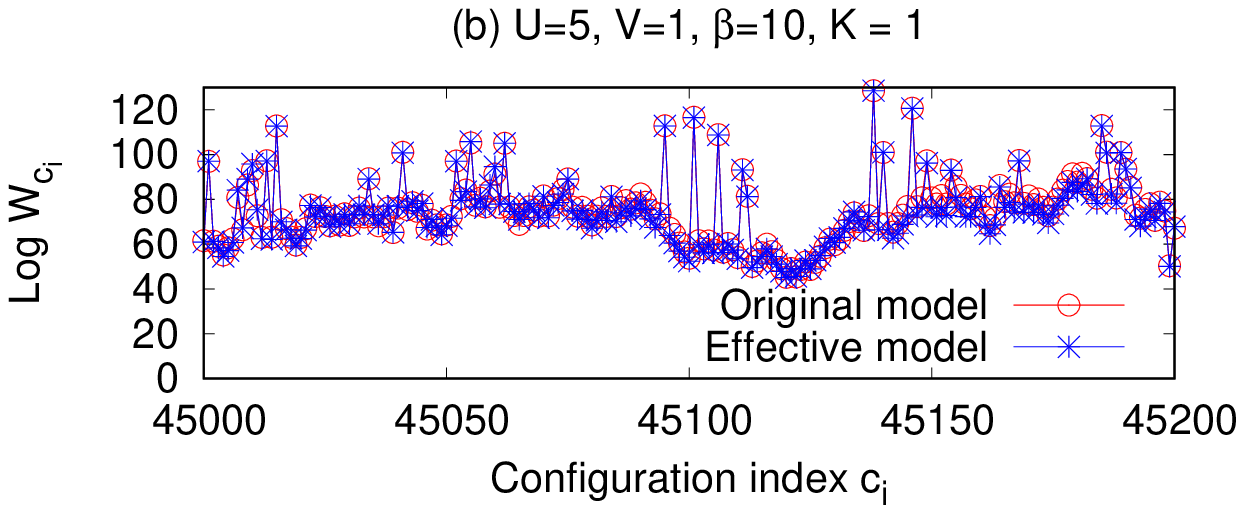}} 
    \end{tabular}
\end{center}
\caption{
(Color online) Comparison between the original and effective weights with $m_{c} = 12$. 
We set $V=1$, $\beta = 10$, and $K = 1$. 
The configurations are generated by the Markov process in the original CTAUX simulation. 
The expansion order $n$ changes in the simulation. 
\label{fig:sfig1}
 }
\end{figure}

To show the efficiency of the trained DGF, we plot the original and effective weights $w_{c_{i}}$ and $w_{c_{i}}^{\rm eff}$. 
We set $m_{s} = 12$ $V=1$, $\beta = 10$, and $K = 1$. 
The configurations are generated by the Markov process in the original CTAUX simulation. 
The expansion order $n$ changes in the simulation. 
As shown in Fig.~\ref{fig:sfig1}, one can clearly find that the weights between these two methods are quite similar.

\section{Local updates in SL-CTAUX}
We show that the calculation cost of the local updates is ${\cal O}(\exn)$. 
We consider the configuration with the expansion order $n$ and the insertion of a vertex with the auxiliary spin $s$ at the imaginary time $\tau$. 
The weight $w_{n+1}^{\rm eff}$ is expressed as 
\begin{align}
\ln w_{n+1}^{\rm eff} &= \frac{1}{n+1} \sum_{i,j}^{n+1}g(\tau_{i}-\tau_{j}) s_{i} s_{j} + 
\frac{1}{n+1} \sum_{i,j}^{n+1}h(\tau_{i}-\tau_{j}) 
+ f(n+1), \\
&= \frac{1}{n+1}  \left(
n w_{n}^{\rm eff} + 2 s \sum_{j=1}^{n} g(\tau - \tau_{j}) s_{j} + g(0) \right. \nonumber \\
&\left. +  2 \sum_{j=1}^{n} h(\tau - \tau_{j})+ h(0)
 \right) + f(n+1).
\end{align}
Thus, the ratio $w_{n+1}^{\rm eff} /w_{n}^{\rm eff}$ is rewritten as 
\begin{align}
\ln \frac{w_{n+1}^{\rm eff}}{w_{n}^{\rm eff}} &= 
\frac{1}{n+1} \left( 
 2 s \sum_{j=1}^{n} g(\tau - \tau_{j}) s_{j} + g(0)
 + 2 \sum_{j=1}^{n} h(\tau - \tau_{j})
  \right. \nonumber \\ 
 &\left. + h(0)
\right) 
- \frac{w_{n}^{\rm eff} }{n+1}  + f(n+1) - f(n).
\end{align}
In the case of removal update, 
The ratio $w_{n-1}^{\rm eff} /w_{n}^{\rm eff}$ becomes 
\begin{align}
\ln \frac{w_{n-1}^{\rm eff}}{w_{n}^{\rm eff}} &= 
\frac{1}{n-1} \left( 
-2 s \sum_{j=1}^{n} g(\tau - \tau_{j}) s_{j} + g(0)
 - 2 \sum_{j=1}^{n} h(\tau - \tau_{j}) \right. \nonumber \\ 
 &\left. + h(0)
\right) 
 + \frac{w_{n}^{\rm eff} }{n-1}  + f(n-1) - f(n).
\end{align}
Thus, the calculation cost of the local updates is ${\cal O}(\exn)$.


\begin{thebibliography}{99}
\bibitem{note2}
By convergence we mean the mean squared error in the linear regression, or the average acceptance rate / autocorrelation time of the cumulative update does not change significantly with $ m_c $. The typical values of $ m_c $ could be roughly estimated by the ``steepness'' of functions $ J(\tau) $ and $ L(\tau) $ near $ \tau=0 $ (Fig.~\ref{fig:fig2}). As will be pointed out later in the paper, these functions will become more localized (steeper) at $  \tau = 0 $ when the temperature is low or when $ U $ is large, in which case more high order Chebyshev polynomials are needed to fit the interaction well.
\bibitem{stab}
Here we briefly comment on the numerical stability in the linear regression. The linear regression suffers from the instability problems either when there is inherent collinearity between the samples or the sample size is very small. Usually in SLMC, the training data could be made sufficiently uncorrelated and could be generated as much as we want by doing MC simulation. Therefore, the instability problem or overfitting is not likely to happen. However, in the scenario when generating training data is costly, this may become an issue. In this case, one may exploit techniques like ridge regression to avoid overfitting.
\end{thebibliography}
\end{document}